\documentstyle[aps,preprint,eqsecnum,tighten,floats,epsf,rotate,prd]{revtex}

\newcommand{\rf}[4]{{\em {#1}} {\bf #2}, #3 (#4)}

\newcommand{\pr}{Phys.\ Rev.\ }
\newcommand{\zf}{Z.\ Phys.\ }
\newcommand{\np}{Nucl.\ Phys.\ }
\newcommand{\beq}{\begin{equation}}
\newcommand{\eeq}{\end{equation}}
\newcommand{\beqa}{\begin{eqnarray}}
\newcommand{\eeqa}{\end{eqnarray}}

\newcommand{\bra}{\langle}
\newcommand{\ket}{\rangle}
\newcommand{\Tr}{{\rm Tr}\,}
\newcommand{\muhat}{\hat{\mu}}
\newcommand{\qhat}{\hat{q}}

\newcommand{\plus}{\makebox[15pt][c]{$+$}}
\newcommand{\minus}{\makebox[15pt][c]{$-$}}
\newcommand{\err}[2]{\raisebox{0.08em}{\scriptsize{$\begin{array}{@{}l@{}}
                          \plus\makebox[0.95em][r]{#1} \\[-0.12em]
                          \minus\makebox[0.95em][r]{#2} 
                        \end{array}$}}}
\newcommand{\er}[2]{\raisebox{0.08em}{\scriptsize{$\begin{array}{@{}l@{}}
                          \plus\makebox[0.55em][r]{#1} \\[-0.12em] 
                          \minus\makebox[0.55em][r]{#2} 
                        \end{array}$}}}

\begin{document}

\preprint{\vbox{\noindent Submitted to 
                          {\it Phys.\ Rev.\ D}
\hfill ADP-98-4/T284\\
\null \hfill LTH 421\\}} 

\title{Gluon Propagator in the Infrared Region}

\author{Derek B.\ Leinweber\footnote{
E-mail:~dleinweb@physics.adelaide.edu.au ~$\bullet$~ Tel:
+61~8~8303~3548 ~$\bullet$~ Fax: +61~8~8303~3551 \hfill\break
\null\quad\quad
WWW:~http://www.physics.adelaide.edu.au/theory/staff/leinweber/}, 
Jon Ivar Skullerud\footnote{
E-mail:~jskuller@physics.adelaide.edu.au ~$\bullet$~ Tel:
+61~8~8303~3426 ~$\bullet$~ Fax: +61~8~8303~3551 \hfill\break
\null\quad\quad
WWW:~http://www.physics.adelaide.edu.au/$^\sim$jskuller/} 
 and Anthony G.\ Williams\footnote{
E-mail:~awilliam@physics.adelaide.edu.au ~$\bullet$~ Tel:
+61~8~8303~3546 ~$\bullet$~ Fax: +61~8~8303~3551 \hfill\break
\null\quad\quad
WWW:~http://www.physics.adelaide.edu.au/theory/staff/williams.html}
}
\address{Special Research Centre for the Subatomic Structure of Matter
and The Department of Physics and Mathematical Physics, University of
Adelaide, Adelaide SA 5005, Australia}
\author{Claudio Parrinello\footnote{
E-mail:~claudio@amtp.liv.ac.uk ~$\bullet$~ Tel:
+44~151~794~3775 ~$\bullet$~ Fax: +44~151~794~3784 \hfill\break
\null\quad\quad
WWW:~http://www.liv.ac.uk/TheoPhys/thep/people/cparrinello.html}
}
\address{Department of Mathematical Sciences, University of Liverpool,
Liverpool L69 3BX, England}
\author{(UKQCD Collaboration)}

\maketitle

\begin{abstract} 
The gluon propagator is calculated in quenched QCD for two different
lattice sizes ($16^3 \times 48$ and $32^3 \times 64$) at $\beta =
6.0$.  The volume dependence of the propagator in Landau gauge is
studied.  The smaller lattice is instrumental in revealing finite
volume and anisotropic lattice artefacts.  Methods for minimising
these artefacts are developed and applied to the larger lattice data.
New structure seen in the infrared region survives these conservative
cuts to the lattice data.  This structure serves to rule out a number
of models that have appeared in the literature.  A fit to a simple
analytical form capturing the momentum dependence of the
nonperturbative gluon propagator is also reported.
\end{abstract}

\vspace{1cm}
\parskip=2mm


\section{Introduction}
\label{sec:intro}

The infrared behaviour of the nonperturbative gluon propagator has
been the subject of extensive research and debate.  Knowledge of this
behaviour is generally regarded as being central to understanding the
mechanism of confinement in quantum chromodynamics (QCD).  Moreover,
it is of great importance in various model-based approaches.  For
example, some studies based on Dyson--Schwinger equations
\cite{mandelstam,bp,bbz} have indicated that an infrared enhanced
gluon propagator may be required for confinement; however, others
\cite{cornwall,gribov,vsmekal} have pointed to the possibility of a
dynamically generated gluon mass, or other forms leading to an
infrared finite (or even vanishing) propagator (see \cite{cdr-agw} and
references therein).

   Computing the gluon propagator directly on the lattice should
provide an opportunity to resolve these contradictory claims.
However, previous lattice studies \cite{bernard,marenzoni} have been
unable to access the ``deep'' infrared region where the most
interesting nonperturbative behaviour is expected.  Significant finite
volume effects introduced through zero momentum components 
prevent the study of the infrared
behaviour of the propagator on a small lattice.

   The main aim of this study is to obtain a definite behaviour for
the gluon propagator for momenta smaller than 1 GeV, where the
interesting physics is expected to reside.  In the following, we will
report such results for momenta as small as 0.4 GeV.  These momenta
are small enough to reveal new structure in the gluon propagator and
to place strong constraints on its infrared behaviour.

\section{The gluon propagator on the lattice}
\label{sec:gluon-lat}

The gauge links $U_\mu(x)$ may be expressed in terms of the continuum
gluon fields as
\beq
U_\mu(x) = {\cal P} e^{ig_0\int_x^{x+\muhat}A_\mu(z)dz} 
= e^{ig_0 aA_\mu(x+\muhat/2)} + {\cal O}(a^3)\, .
\eeq
where ${\cal P}$ denotes path ordering.
From this, the dimensionless lattice gluon field $A^L_{\mu}(x)$ may be
obtained via
\beq
A^L_\mu(x+\muhat/2) = \frac{1}{2ig_0}\left(U_\mu(x)-U^{\dagger}_\mu(x)\right)
 - \frac{1}{6ig_0}\Tr\left(U_\mu(x)-U^{\dagger}_\mu(x)\right) \, ,
\eeq
which is accurate to ${\cal O}(a^2)$. Denoting the discrete momenta
available on a finite, periodic volume by $\qhat$, the momentum space gluon
field is
\beqa
A_\mu(\qhat) & \equiv & \sum_x e^{-i\qhat\cdot(x+\muhat/2)}
 A^L_\mu(x+\muhat/2) \nonumber \\
 & = & \frac{e^{-i\qhat_{\mu}a/2}}{2ig_0}\left[\left(U_\mu(\qhat)-U^{\dagger}_\mu(-\qhat)\right)
 - \frac{1}{3}\Tr\left(U_\mu(\qhat)-U^{\dagger}_\mu(-\qhat)\right)\right] , 
\eeqa
where $U_\mu(\qhat)\equiv\sum_x e^{-i\qhat x}U_\mu(x)$ and
$A_\mu(\qhat)\equiv t^a A_{\mu}^a(\qhat)$.  The available momentum
values $\qhat$ are given by
\beq
\qhat_\mu  = 
\frac{2 \pi n_\mu}{a L_\mu}, \qquad
n_\mu=0,\ldots,L_\mu-1
\eeq
where $L_\mu$ is the length of the box in the $\mu$ direction.  The
factor $e^{-i\qhat_{\mu}a/2}$, which comes from the gauge fields being
defined on the links rather than the sites of the lattice, is crucial
to removing ${\cal O}(a)$-errors and in particular obtaining the
correct tensor structure for the gluon propagator \cite{ggg}.  The
gluon propagator $D_{\mu\nu}^{ab}(\qhat)$ is defined by
\beq
D^{ab}_{\mu\nu}(\qhat) \equiv \bra A_\mu^a(\qhat)A_\nu^b(-\qhat)\ket \, .
\eeq

We choose to study the gauge dependent propagator in the 
Landau gauge, which 
can be implemented numerically by maximising 
$F [g] = \sum_{\mu,x}\Re\Tr
U^g_\mu(x)$, where
\beq
U^g_\mu(x) = g(x)U_\mu(x)g^{\dag}(x+\muhat).
\eeq 
In the continuum limit, this is related to the
fact that fields satisfying the Landau gauge condition correspond to
stationary points of $F^{cont} [g] = \parallel\!  A^g\!\parallel^2 =
\int d^4\!x \Tr (A^g_\mu)^2 (x)$ \cite{gribov}.

The Landau gauge gluon propagator in the continuum has the form
\beq
D_{\mu\nu}^{ab}(q) =
(\delta_{\mu\nu}-\frac{q_{\mu}q_{\nu}}{q^2})\delta^{ab}D(q^2)
\, ,
\label{eq:landau_prop}
\eeq
The scalar function $D(q^2)$ can then be extracted using
\beq
D(q^2) = \frac{1}{3}\sum_{\mu}\frac{1}{8}\sum_{a}D_{\mu\mu}^{aa}(q).
\label{eq:scalar-prop}
\eeq
For $a \rightarrow 0$, 
the lattice propagator is related to the continuum one by 
$a^2 D(q^2) = D^{\rm cont}(q^2) + {\cal O}(a^2)\,$.

A well-known lattice artefact is that the tree level propagator of a
massless boson field does not reproduce the expected continuum result of
\beq
D^{(0)}(q^2) = \frac{1}{q^2} ,
\label{eq:tree}
\eeq
but rather produces
\beq
a^2 D^{(0)}(\qhat) = \frac{a^2}{\sum_{\mu}(2\sin \qhat_\mu a/2)^2} \, .
\label{eq:lat_tree}
\eeq
In the following, we are particularly interested in the quantity 
$q^2 D(q^2)$, which is expected to approach 1 up to logarithmic
corrections as $q^2\to\infty$.
To ensure this result we will work with the momentum variable
defined as\footnote{The momenta $q$ and $\qhat$ are often
defined the other way around in the lattice literature.  However, we feel it
is more instructive to define $q$ as above, such that
lattice results reproduce the
continuum formula (\protect{\ref{eq:landau_prop}}) and the tree level
formula (\protect{\ref{eq:tree}}).}
\beq
q_{\mu} \equiv \frac{2}{a} \sin\frac{\qhat_{\mu} a}{2}\, .
\label{eq:latt_momenta}
\eeq

\section{Simulation parameters}

We have analysed 75 configurations at $\beta=6.0$, on a $32^3\times
64$ lattice.  Using the value of $a^{-1}=1.885$ GeV based on the
string tension in \cite{bs},
this corresponds to a physical volume of ($3.35^3\times 6.70$)fm.  For
comparison, we have also studied an ensemble of 125 configurations on
a smaller volume of $16^3\times 48$, with the same lattice spacing.  

The gauge configurations were generated using a combination of 
seven over-relaxation and one Cabibbo--Marinari updates, with a separation
between configurations of 1000 combined updates for the large lattice
and 800 for the smaller lattice.
Both lattices were fixed to Landau gauge using a Fourier accelerated
steepest descent 
algorithm \cite{cthd}.  An accuracy of
$\frac{1}{VN_C}\sum_{\mu,x}|\partial_{\mu}A_{\mu}|^2 <10^{-12}$ was
achieved.\footnote{The average link trace 
$\frac{1}{3}\bra\Re\Tr U_\mu\ket$ for the
gauge-fixed configurations at $\beta=6.0$ was determined to 0.8609(7)
for the small lattice and 0.8617(1) for the large lattice.}

\section{Finite Size Effects and Anisotropic Behaviour}
\label{sec:artefacts}

\subsection{Small Lattice Analysis}

\begin{figure}[t]
\begin{center}
\epsfysize=11.6truecm
\leavevmode
\rotate[l]{\vbox{\epsfbox{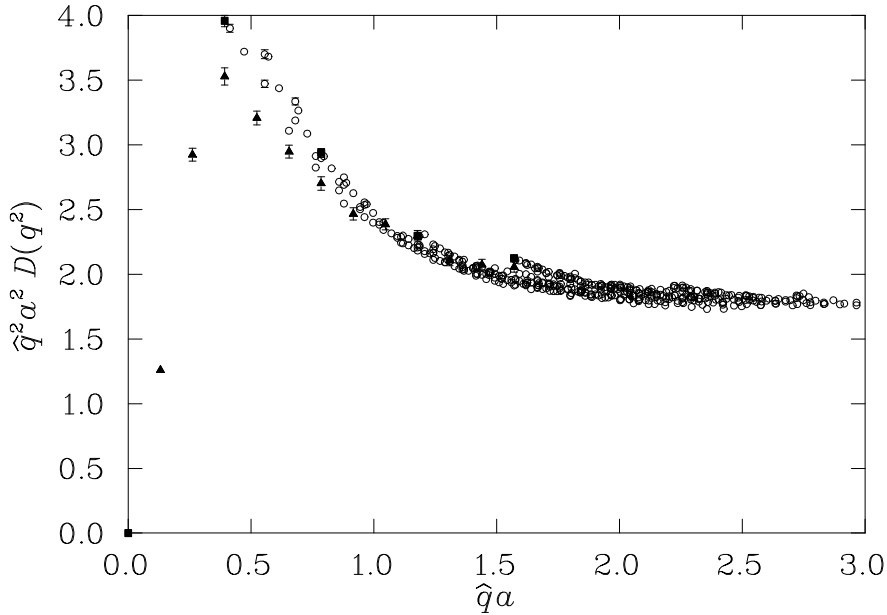}}}
\end{center}
\caption{The gluon propagator from our small lattice multiplied by
$\qhat^2 a^2$ plotted as a function of momenta $\qhat a$.  Values for each
momentum direction are plotted separately.  Only a $Z_3$ averaging has been
performed.  Filled squares denote momenta directed along spatial
axes, while filled triangles denote momenta directed
along the time axis.  Other momenta are indicated by open circles.}
\label{fig:small_qhat}
\end{figure}

We begin by considering the effect of the kinematic correction
introduced through the change of variables in (\ref{eq:latt_momenta}).
In the absence of this correction, data in the high momentum region
are expected to display significant anisotropy when shown as a
function of $\qhat$.  This is confirmed in Fig.~\ref{fig:small_qhat},
which shows the gluon propagator multiplied by $\qhat^2 a^2$ and
plotted as a function of $\qhat a$.  Here and in the following, a
$Z_3$ averaging is performed on the data, where for example the
momentum along (t,x,y,z) = (1,2,1,1) is averaged with (1,1,2,1) and
(1,1,1,2).

\begin{figure}[t]
\begin{center}
\epsfysize=11.6truecm
\leavevmode
\rotate[l]{\vbox{\epsfbox{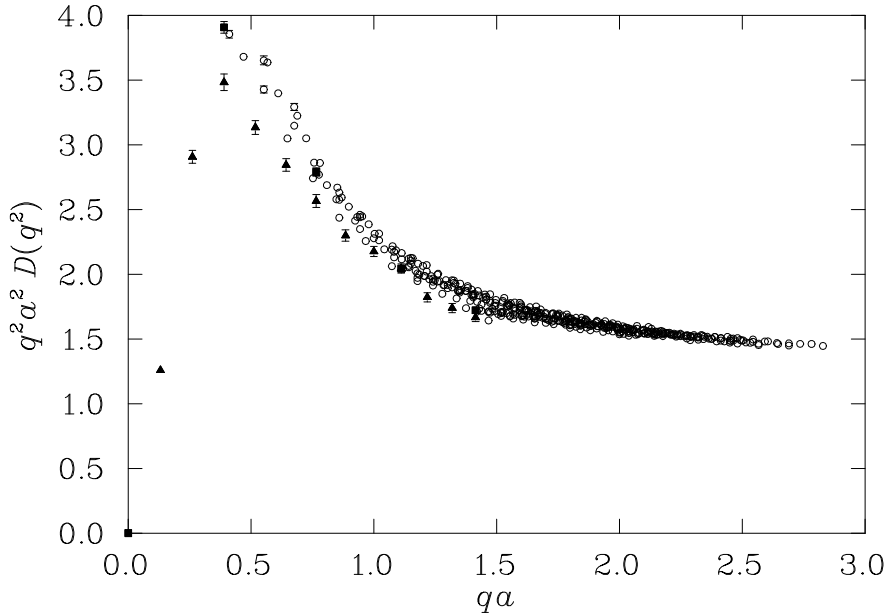}}}
\end{center}
\caption{The gluon propagator from our small lattice multiplied by
$q^2 a^2$ plotted as a function of momenta $qa$.  Values for each
momentum direction are plotted separately.  Only a $Z_3$ averaging has
been performed.  The symbols are as in Fig.~\protect{\ref{fig:small_qhat}}.
Finite volume errors are expected to be largest for the purely
time-like momenta, as the three shorter spatial directions have
momentum components equal to zero and hence the effects of the nearby
spatial boundaries are expected to be at a maximum.}
\label{fig:small_comp}
\end{figure}

We expect the kinematic correction to reduce anisotropy, particularly
in the large momentum region.  Fig.~\ref{fig:small_comp} shows the
gluon propagator multiplied by the factor $q^2 a^2$ and plotted as a
function of $q a$ for momenta directed throughout the lattice.  The
anticipated reduction of anisotropy for $q a > 1.5$ is nicely
displayed in this figure.  A similar result was found in
\cite{marenzoni}.

Since the low momentum region holds the greatest nonperturbative
interest, it is instructive to stress that the low momentum points
displayed in Figs.~\ref{fig:small_qhat} and \ref{fig:small_comp} are
insensitive to the whether one plots as a function of $q$ or $\qhat$.
It is also useful to note that on any finite lattice, $D(q^2)$ will be
finite at $q^2 a^2=0$. Hence any lattice calculation must give $q^2
a^2 D(q^2)$ vanishing at $q^2 = 0$.

On the small lattice, we also see significant anisotropy in the data
which have their origin in finite volume artefacts.  Finite size
effects are expected to be largest where the momentum component along
one or more of the lattice axes is zero.  Since the length of the
lattice in the time direction is three times that of the spatial
directions, we notice a clear difference between points which
correspond to momenta directed along spatial axes from those purely in
the time direction.  These finite volume artefacts are clearly
displayed at small momenta by the discrepancies between the filled
squares (denoting momenta directed along spatial axes), and the filled
triangles (denoting momenta directed along the time axis).

Some residual anisotropy remains at moderate momenta around $q a \sim
1.5$ despite including the kinematical correction of
(\ref{eq:latt_momenta}).  This anisotropy is clearly displayed in
Fig.~\ref{fig:small_comp} by the filled squares and triangles denoting
momenta directed along lattice axes lying below the majority of points
from off-axis momenta\footnote{In Fig.~\protect\ref{fig:small_comp} it
appears that this anisotropy disappears as one goes to even larger
momenta.  However, this is not necessarily the case.  Only momentum
components up to $\qhat_\mu a = 4 \times 2\pi/16$ have been selected
in preparing Fig.~\protect\ref{fig:small_comp}.  This means that the
largest momenta are not accessed through any single Cartesian
direction.} for $q a \sim 1.4$.  Since tree-level and one-loop $O(4)$
breaking effects should be removed by the kinematical correction, the
remaining anisotropy appears to have its origin in quantum effects
beyond one loop.

\begin{figure}[p]
\begin{center}
\epsfysize=11.6truecm
\leavevmode
\rotate[l]{\vbox{\epsfbox{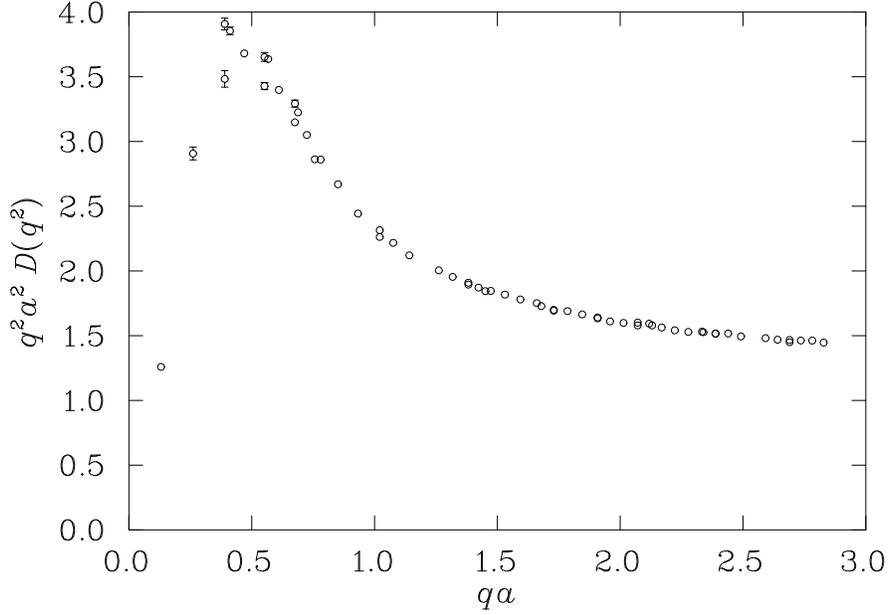}}}
\end{center}
\caption{The gluon propagator from our smaller lattice multiplied by
$q^2 a^2$.  The points displayed in this plot lie within a cylinder of
radius $\Delta\qhat a < 1\!\times\!2\pi/16$ directed along the diagonal 
$(t,x,y,z) = (1,1,1,1)$ of the lattice.}
\label{fig:small_comp_cyl1}
\end{figure}

In order to minimise lattice artefacts for large momentum components
we select momentum vectors lying within a cylinder directed along the
diagonal $(t,x,y,z) = (1,1,1,1)$ of the lattice.  This allows one to
access the largest of momenta with the smallest of components.  We
found the selection of a cylinder with a radius of one spatial
momentum unit ($\Delta\qhat a < 1\!  \times\!2\pi/L_s$, where $L_s$ is
the number of sites along a spatial axis) provides a reasonable number
of points falling along a single curve for large momenta.  The data
surviving this cut are displayed in Fig.~\ref{fig:small_comp_cyl1}.

\begin{figure}[p]
\begin{center}
\epsfysize=11.6truecm
\leavevmode
\rotate[l]{\vbox{\epsfbox{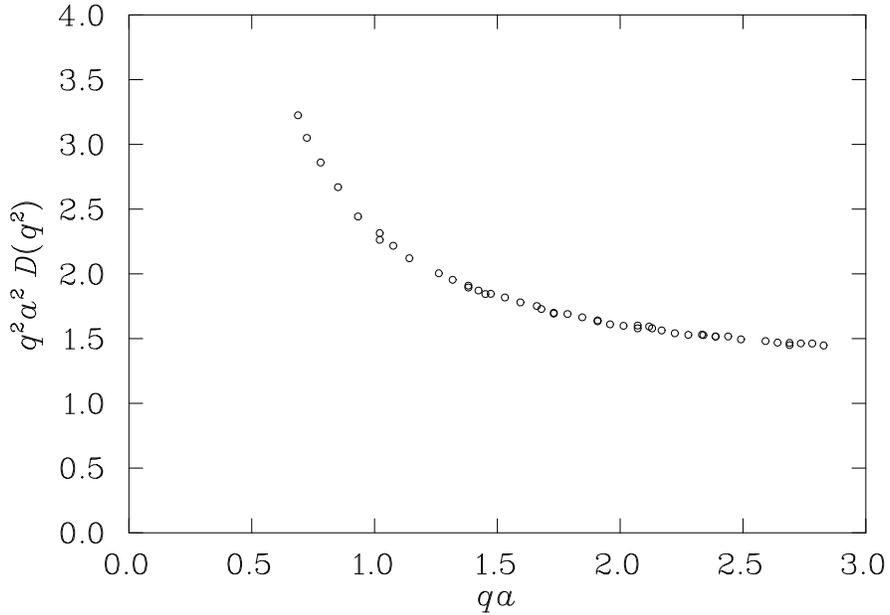}}}
\end{center}
\caption{The gluon propagator from our smaller lattice multiplied by
$q^2 a^2$.  The points displayed in this plot lie within a cylinder of
radius $\Delta\qhat a < 1\!\times\!2\pi/16$ directed along the
diagonal of the lattice and within a cone of $20^\circ$ measured from
the diagonal at the origin.}
\label{fig:small_comp_ang20_cyl1}
\end{figure}

However, this cut does not address the large finite volume errors
surviving in Fig.~\ref{fig:small_comp_cyl1}.  To remove these
problematic momenta, we consider a further cut designed to remove
momentum vectors which have one or more vanishing components.  This is
implemented by keeping only momentum directions that lie within a
certain solid angle of the diagonal.  We found that a cone of angle
$20^\circ$ measured from the diagonal at the origin was sufficient to
provide a set of points lying along a smooth curve.
Fig.~\ref{fig:small_comp_ang20_cyl1} displays these data.  After these
conservative cuts, there is relatively little structure left in the
infrared region on our small lattice.  Armed with this knowledge of
how to obtain reliable lattice data, we now turn our attention to the
gluon propagator data obtained from our larger lattice.

\subsection{Large Lattice Analysis}

\begin{figure}[t]
\begin{center}
\epsfysize=11.6truecm
\leavevmode
\rotate[l]{\vbox{\epsfbox{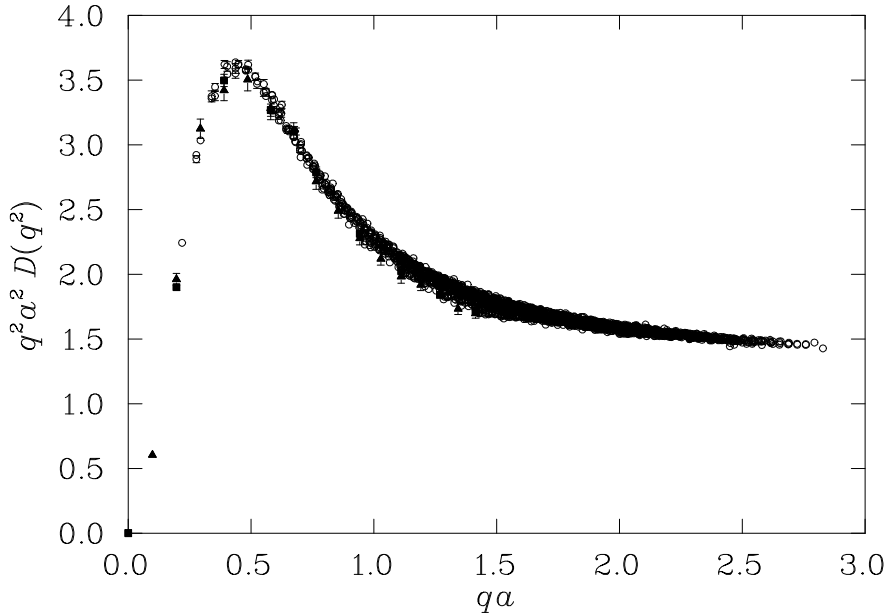}}}
\end{center}
\caption{The gluon propagator from our larger lattice multiplied by
$q^2 a^2$ plotted as a function of momenta $qa$.  Values for each
momentum direction are plotted separately.  Only a $Z_3$ averaging has
been performed for the data shown in this figure.  Plotting symbols
are as in Fig.~\protect\ref{fig:small_comp}.  Finite volume errors are
greatly reduced compared to the results from the smaller lattice, as
displayed by the overlap of points obtained from spatial and time-like
momentum vectors.  However, significant anisotropy is apparent for
larger momenta.}
\label{fig:large_comp}
\end{figure}

Fig.~\ref{fig:large_comp} displays the gluon propagator data for all
momentum directions and values on the larger lattice.  Again, only a
$Z_3$ averaging has been performed.  Examination of the infrared
region indicates that finite volume artefacts are very small on the
larger lattice.  In particular, the agreement between purely spatial
(filled squares) and time-like momentum vectors (filled triangles) at
$qa = 0.20$ appears to indicate that finite size effects are
relatively small on this lattice.  Such an observation is consistent
with topological studies of the QCD vacuum which provide some insight
into the typical scale of QCD vacuum fluctuations.

\begin{figure}[p]
\begin{center}
\epsfysize=11.6truecm
\leavevmode
\rotate{\vbox{\epsfbox{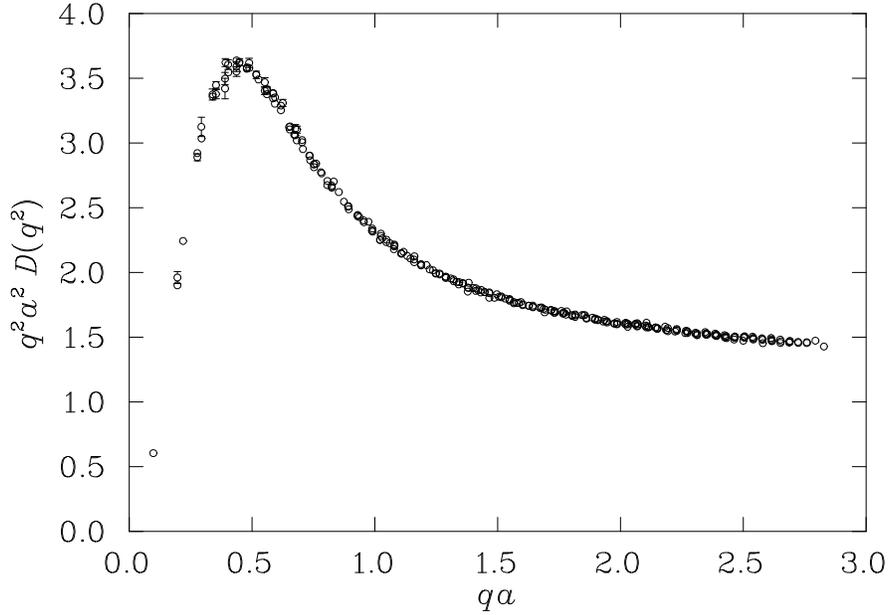}}}
\end{center}
\caption{The gluon propagator from our larger lattice multiplied by
$q^2 a^2$.  The points displayed in this plot lie within a cylinder of
radius $\Delta\qhat a < 2\!\times\!2\pi/32$ directed along the
diagonal of the lattice.  The first data point will be ignored in all
fits, since it is not possible to assess the finite size effects for
this point.  The agreement between the next pair of data points
indicates that finite size effects here are small.}
\label{fig:large_comp_cyl2}
\end{figure}

At large momenta, $q a > 1.0$, significant anisotropy is observed,
similar to those displayed in Fig.~\ref{fig:small_comp}.  The fact
that this anisotropy occurs at the same momentum values and with the
same magnitude on both lattices confirms our previous argument that
they result from finite lattice spacing errors as opposed to finite
volume errors.  Similar behaviour is expected in this region as the
lattice action and lattice spacing are the same for our two lattices.
To side-step this problem, we adopt the same cut as before.  On this
larger lattice, all momenta must lie within a cylinder of radius two
spatial momentum units directed along the lattice diagonal.
Fig.~\ref{fig:large_comp_cyl2} displays the momenta surviving this
cut.

\begin{figure}[p]
\begin{center}
\epsfysize=11.6truecm
\leavevmode
\rotate{\vbox{\epsfbox{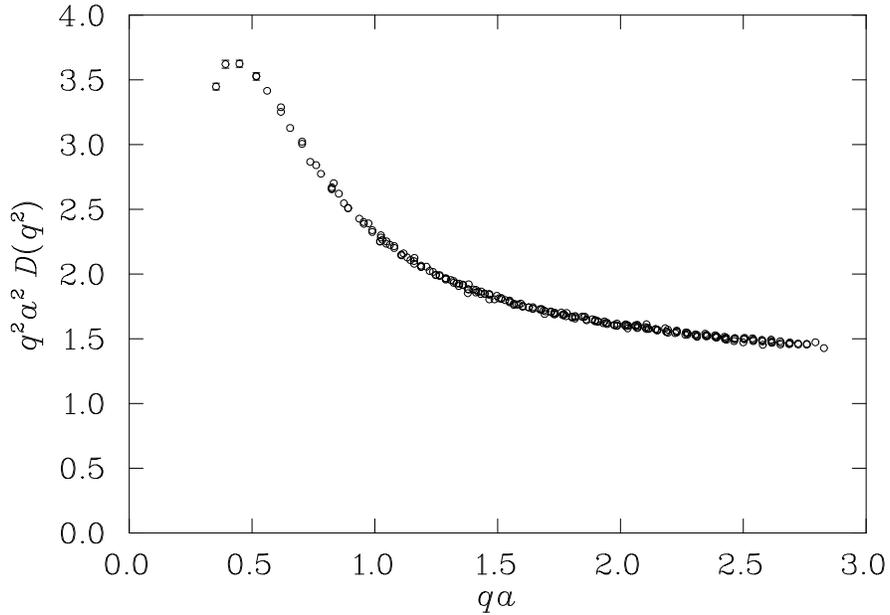}}}
\end{center}
\caption{The gluon propagator from our larger lattice multiplied by
$q^2 a^2$.  The points displayed in this plot lie within a cylinder of
radius $\Delta\qhat a < 2\!\times\!2\pi/32$ directed along the
diagonal of the lattice and within a cone of $20^\circ$ measured from
the diagonal at the origin.}
\label{fig:large_comp_ang20_cyl2}
\end{figure}

While the two points at $qa=0.20$ indicate that finite volume errors
are small on this lattice, we also consider the additional angular
cut, requiring that all points must lie within a cone of opening angle
$20^\circ$ from the diagonal at the origin.  The data surviving this
cautiously conservative cut are illustrated in
Fig.~\ref{fig:large_comp_ang20_cyl2}.  It is interesting to see that
the turnover in $q^2a^2 \, D(q^2)$ in the infrared region survives
even this extreme cut on our larger lattice.

\section{Modelling the propagator}
\label{sec:numbers}

The approach we have described here differs in several respects from
that of \cite{marenzoni}.  In particular, the momentum vectors
surviving our cuts are very different.  In \cite{marenzoni} only a
small number of low-lying spatial momentum values are used along with
all available time momentum values.  In contrast, we have treated
momenta in the spatial and time directions on an equal footing and
selected momenta lying near the 4-dimensional diagonal, where lattice
artefacts are expected to be minimal.  In fact, nearly all the momenta
used in \cite{marenzoni} would be excluded by our cuts to the data.

In addition, we include the kinematical correction of
(\ref{eq:latt_momenta}).  The authors of Ref.\ \cite{marenzoni} do not
include such a correction when performing their fits.  Instead, their
fits are constrained to the low-momentum region where such artefacts
are hoped to be small.

We have considered a number of models for the propagator which have
been suggested in the literature, as well as 
some simple analytical forms which are intended to capture the essence
of the nonperturbative gluon propagator.  A detailed analysis of these
models is currently in progress \cite{next}.

The data for the fit are those obtained on the large lattice with the
cylindrical cut.  To balance the sensitivity of the fit over the
available range of $qa$, we average adjacent lattice momenta lying
within $\Delta qa < 0.005$.  In all fits the first point, at $qa \sim
0.1$, is omitted as it may be sensitive to the finite volume of the
lattice.

\begin{figure}[t]
\begin{center}
\epsfysize=11.6truecm
\leavevmode
\rotate{\vbox{\epsfbox{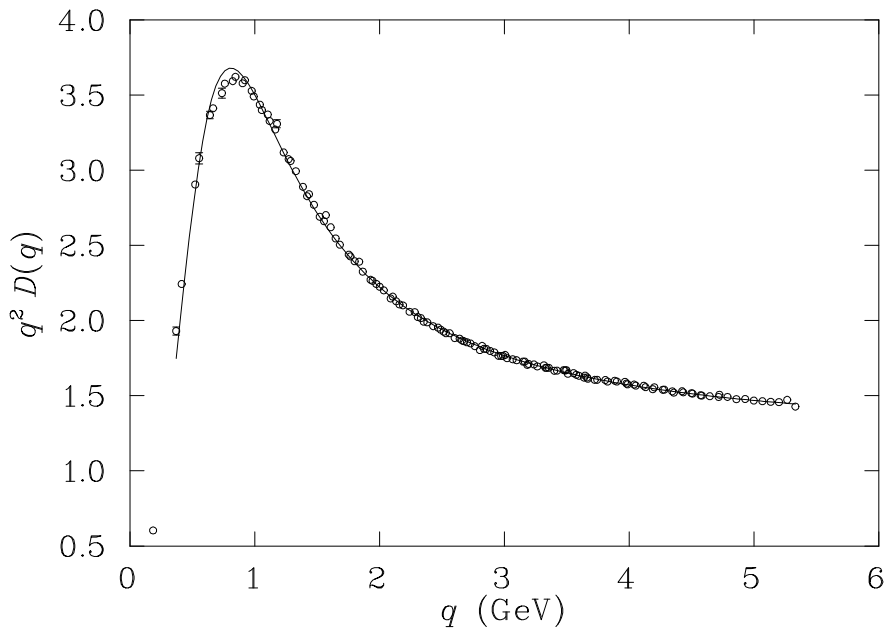}}}
\end{center}
\caption{The gluon propagator multiplied by $q^2$, with nearby
points averaged.  The line illustrates our best fit to the form
defined in Eq.~(\protect{\ref{eq:model}}).  The fit is performed over
all points shown, excluding the one at the lowest momentum value, which
may be sensitive to the finite volume of the lattice.}
\label{fig:fit_large_model4}
\end{figure}

\begin{figure}[t]
\begin{center}
\epsfysize=11.6truecm
\leavevmode
\rotate{\vbox{\epsfbox{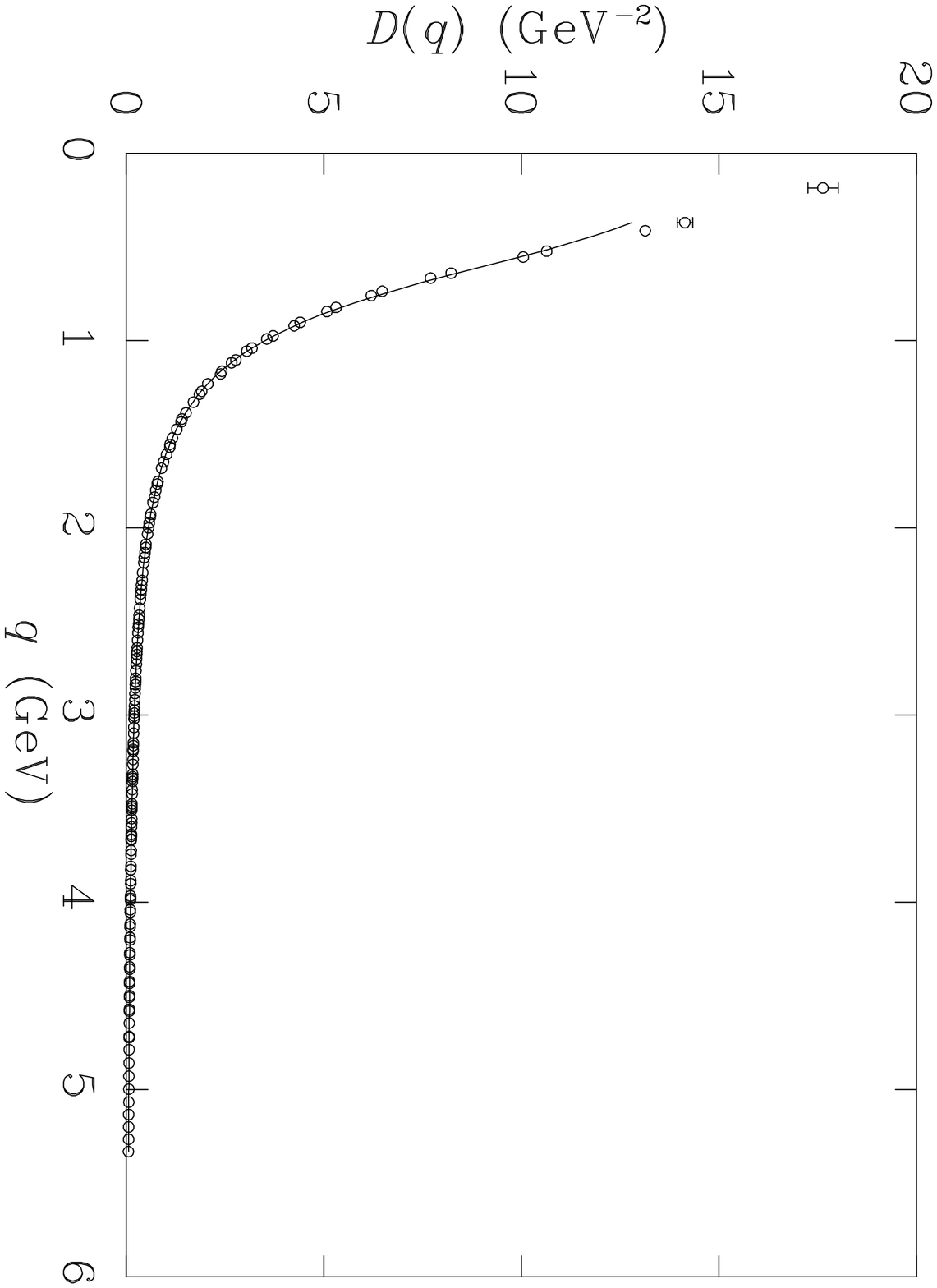}}}
\end{center}
\caption{The gluon propagator in physical units,
with nearby
points averaged.  The line illustrates our best fit to the form
defined in Eq.~(\protect{\ref{eq:model}}).  The fit is the same as
that shown in Fig.~\protect{\ref{fig:fit_large_model4}}.
The scale is taken from the value for the string tension quoted in
\protect{\cite{bs}}. 
}
\label{fig:gluon_phys_large_model4}
\end{figure}

Our results so far indicate that the following analytical
form
\beq
D(q^2) = Z\left(\frac{A}{(q^2 a^2)^{1+\alpha}+(M^2)^{1+\alpha}} +
\frac{1}{q^2 a^2+M^2} \right) \, ,
\label{eq:model}
\eeq
provides a satisfactory description of the data over a wide
range of momenta.
Our best fit 
to (\ref{eq:model}) is illustrated in
Figs.~\ref{fig:fit_large_model4} and
\ref{fig:gluon_phys_large_model4}.
This fit yields $\chi^2/{\rm
dof} = 3.5$, a somewhat high value.  However, if the first
four points are omitted, a more acceptable value of 1.6 is found.

We have studied the stability of the fits by varying the fitting
range, and 
formula (\ref{eq:model}) turns out to give stable values for
all the parameters over a wide region of fitting ranges.  Our best
estimates for the parameters, using all the available data, are
\beqa
Z & = & 1.214\er{5}{5}\err{70}{25} \\
A & = & 1.059\er{9}{9}\err{45}{65} \\
\alpha & = & 0.784\er{4}{4}\err{80}{20} \\
M & = & 0.375\er{2}{2}\err{50}{10} \, ,
\eeqa
where the errors denote the uncertainties in the last digit(s) of the
parameters.  
The first set of errors are statistical uncertainties in
the fit to the full data set, and 
the second set of
errors are due to fluctuations in the parameters as the
fitting interval is varied.  The estimate for $M$
corresponds to a physical value for the parameter in the region of 700 MeV.

\section{Conclusions}
\label{sec:conclusions}

We have performed an accurate, nonperturbative study of the infrared
behaviour of the gluon propagator in the pure gauge theory on the
lattice.  We were able to isolate a set of data points for which
systematic lattice errors seem to be negligible.  Our data indicate an
infrared behaviour less singular than $1/q^2$.  This can be inferred
from our plots by noticing the clear turnover in the behaviour of $q^2
a^2 \, D(q)$ around $q^2 = 1\ {\rm GeV}^2$.  This is in agreement with
suggestions from previous lattice results.
In particular, there is evidence that the nonperturbative gluon
propagator may be infrared finite.  Our data appear to rule out models
like those proposed in \cite{bbz}, which lead to an infrared enhanced
propagator.

Work in progress focuses on a detailed analysis of various analytic
forms for the gluon propagator \cite{next}.  A stability analysis of
the fit parameters is central to identifying the model best able to
describe the gluon propagator in both the nonperturbative and the well
known perturbative regimes.  We are also exploring the possibility of
extracting the value of $D(q=0)$ in the infinite volume limit, which
one may be able to extrapolate from data at different volumes.  A study
of the effects of Gribov copies \cite{cucchieri} and a complete
analysis of the tensor structure of the gluon propagator are also
issues under consideration.

Finally, one promising line of research appears to be the study of the
gluon propagator using improved lattice actions \cite{improve}.  These
should yield a significant reduction of finite lattice spacing
effects.  Then, by performing simulations at larger values of the
lattice spacing, one would be able to measure the propagator on larger
physical volumes, thus gaining access to very low momentum values.
For example, one may be able to use lattice spacings as large as 0.4
fm, so that a modest $16^4$ lattice would have a physical size of 6.4
fm.

\section*{Acknowledgments} 

The numerical work was mainly performed on a Cray T3D based at EPCC,
University of Edinburgh, using UKQCD Collaboration CPU time under 
PPARC Grant GR/K41663. Also, CP acknowledges further support from 
PPARC through an Advanced Fellowship. 
Financial support from the Australian Research Council 
is also gratefully
acknowledged.


\begin{thebibliography}{99}

\bibitem{mandelstam} S.~Mandelstam, \rf{\pr}{D 20}{3223}{1979}

\bibitem{bp} N.~Brown and M.R.~Pennington, \rf{\pr}{D 39}{2723}{1989}

\bibitem{bbz} M.~Baker, J.S.~Ball, F.~Zachariassen, \rf{\np}{B 186}{531}{1981}

\bibitem{cornwall} J.~Cornwall, \rf{\pr}{D 26}{1453}{1982}

\bibitem{gribov} V.N.~Gribov, \rf{\np}{B 139}{19}{1978}; D.~Zwanziger,
\rf{\np}{B 378}{525}{1992};
U.~Habel et al, \rf{\zf}{A 336}{423}{1990} 

\bibitem{vsmekal} L.~von Smekal, A.~Hauck, R.~Alkofer,
\rf{\prl}{79}{3591}{1997}; D.~Atkinson and J.C.R.~Bloch,
hep-ph/9712459 and hep-ph/9802239

\bibitem{cdr-agw} C.D.~Roberts and A.G.~Williams, \rf{Progress in
Particle and Nuclear Physics}{33}{477-575}{1994}

\bibitem{bernard} C.~Bernard, C.~Parrinello, A.~Soni,
\rf{\pr}{D49}{1585}{1994} 

\bibitem{marenzoni} P.~Marenzoni, G.~Martinelli, N.~Stella, 
\rf{\np}{B 455}{339}{1995}; P.~Marenzoni, G.~Martinelli, N.~Stella,
M.~Testa, \rf{\pl}{B 318}{511}{1993}

\bibitem{ggg} B.~All\'es et al, \rf{\np}{B 502}{325}{1997}

\bibitem{bs} G.S.~Bali and K.~Schilling, \rf{\pr}{D 47}{661}{1993}

\bibitem{cthd} C.T.H.~Davies et al, \rf{\pr}{D 37}{1581}{1988}

\bibitem{next} D.B.~Leinweber, C.~Parrinello,
J.I.~Skullerud, A.G.~Williams, in preparation.

\bibitem{cucchieri} 
E.~Marinari, C.~Parrinello, R.~Ricci,  \rf{\np}{B 362}{487}{1991};  
A.~Cucchieri, \rf{\np}{B 508}{353}{1997}

\bibitem{improve} K.~Symanzik, \rf{\np}{B 226}{187}{1983};
M.~Alford et al, \rf{\pl}{B 361}{87}{1995}

\end{thebibliography}
\end{document}